\documentclass[%
 reprint,
 amsmath,amssymb,
 aps,
]{revtex4-2}

\usepackage{multirow}
\usepackage{graphicx}
\usepackage{dcolumn}
\usepackage{bm}
\usepackage{upgreek}
\usepackage{hyperref}

\begin{document}

\preprint{APS/123-QED}

\title{Photoacoustic model for laser-induced acoustic desorption of nanoparticles}

\author{Matthew Edmonds}
\author{James Bateman}%
 \email{j.e.bateman@swansea.ac.uk}
\affiliation{%
  Department of Physics\\
  Faculty of Science and Engineering\\
  Swansea University\\
  SA2 8PP, UK
}%

\date{\today}

\begin{abstract}
Laser-induced acoustic desorption (LIAD) enables loading nanoparticles into optical traps under vacuum for levitated optomechanics experiments. Current LIAD systems rely on empirical optimization using available laboratory lasers rather than systematic theoretical design, resulting in large systems incompatible with portable or space-based applications. We develop a theoretical framework using the photoacoustic wave equation to model acoustic wave generation and propagation in metal substrates, enabling systematic optimization of laser parameters. The model identifies key scaling relationships: surface acceleration scales as $\tau^{-2}$ with pulse duration, while acoustic diffraction sets fundamental limits on optimal spot size $w \gtrsim \sqrt{v\tau d}$. Material figures of merit combine thermal expansion and optical absorption properties, suggesting alternatives to traditional aluminum substrates. The framework validates well against experimental data and demonstrates that compact laser systems with sub-nanosecond pulse durations can achieve performance competitive with existing laboratory-scale implementations despite orders-of-magnitude lower pulse energies. This enables rational design of minimal LIAD systems for practical applications.

\end{abstract}

\maketitle

\section{Introduction}

Levitated optomechanics uses optical forces to trap microscopic objects in vacuum, where the mechanical motion, through isolation from the enviornment, achieves exceptional mechanical quality factors~\cite{gonzalezballestero2021levitodynamics,winstone2023optomechanics}.
Recent demonstrations of real-time optimal state estimation have enabled quantum limited control at room temperature~\cite{setter2018realtime,magrini2021realtime}.
Decoupling from environmental noise enables precision sensing applications including force and acceleration measurements~\cite{ranjit2016zeptonewton,monteiro2020force}.
As sensors, these show promise for searches for new physics, probing short-range deviations from Newtonian gravity, and for high-frequency gravitational waves~\cite{geraci2010short,arvanitaki2013detecting,aggarwal2022searching}.

Proposals for pressure sensing through collision-resolved detection~\cite{barker2024collisionresolved} are supported by recent demonstrations of individual momentum transfer measurements near the standard quantum limit~\cite{tseng2025search}.
Levitated optomechanics may also facilitate tests of quantum mechanics using mesoscopic superpositions~\cite{romero2010quantum,bateman2014nearfield,bose2025massive}, complementing recent free-flight demonstrations~\cite{pedalino2025probing}.

A critical challenge is loading nanoparticles into optical traps under vacuum.
Traditional methods work at ambient pressure using liquid dispersions~\cite{summers2008trapping}, introducing water or solvents that makes operating at high vacuum challenging.
Approaches using piezoelectric launchers~\cite{weisman2022apparatus,khodaee2022dry} can operate at reduced pressures but launch particles over a comparatively large area.
Custom micro-electromechanical systems (MEMS) provide controlled particle launching from individually placed particles~\cite{khorshad2025invacuum}, though current devices achieve minimum sizes of $\sim$4~$\mu$m and have finite particle reservoirs, whereas many applications require sub-micron particles and extended operation made possible by large coated areas.
Hybrid electro-optical approaches~\cite{goldwater2019levitated,bonvin2024hybrid} represent an emerging direction and provide an efficient mechanism to capture often fast-moving ejected particles, but still require a vacuum-compatible loading mechanism.

Laser-induced acoustic desorption (LIAD) offers a promising solution. Originally developed for mass spectrometry~\cite{dow2012laser}, LIAD uses pulsed lasers to generate acoustic waves in metal substrates, creating sufficient surface accelerations on the distant, particle-laden side, as to overcome the surface adhesion forces and eject them into the gas phase. This ejection occurs over a small region, with reproducible velocity and low-jitter timing.

Current LIAD implementations rely on empirical optimization rather than systematic design~\cite{nikkhou2021direct,millen2020optomechanics}. Typical setups use frequency-doubled Nd:YAG lasers (532~nm, 5--25~mJ, 3--5~ns), parameters of which are set by availability and precedent rather than theoretical optimization.

These systems require substantial power supplies and cooling, resulting in devices weighing tens of kilograms which is incompatible with portable accelerometers or compact sensors. Space applications are particularly challenging as small satellites require minimal size, weight, and power while maintaining reliable performance.
Space-based levitated optomechanics missions~\cite{homans2025macroscopic,gajewski2025levitas} require particle loading mechanisms compatible with the space environment's constraints on power, vacuum integrity, and long-term autonomous operation.

This necessitates a transition from empirical selection to systematic optimization based on theoretical understanding. Rather than adapting available laboratory lasers, practical applications require compact systems designed for optimal LIAD performance.

This work develops a theoretical framework for LIAD optimization. We model photoacoustic wave generation using the scalar wave equation with realistic parameters. The approach captures the essential physics of wave generation and propagation while remaining numerical tractable. Theoretical predictions agree well with experimental data, validating the model for system design.

\section{Theory}

\subsection{Photoacoustic wave equation}

The photoacoustic equation models sound waves generated by optical heating, and is valid for laser pulses with durations long compared to the electron-phonon coupling time ($\sim$10~ps)~\cite{allen1987theory,grimvall1981electron} but short compared to the thermal relaxation time ($\sim$1~$\upmu$s)~\cite{xu2006photoacoustic}. This regime encompasses most pulsed laser systems used in LIAD applications, including nanosecond-duration frequency-doubled Nd:YAG lasers commonly employed.

The thermal relaxation time arises from thermal confinement, which requires negligible heat diffusion during the pulse, and imposes the constraint $\tau \ll w^2/D$, where $D \approx 0.1$~$\upmu$m$^2$/ns is a typical thermal diffusivity for metals.
This is well satisfied for most LIAD systems; for example, when focused to $w\sim 10$~$\upmu$m pulses must be shorter than $1$~$\upmu$s. For tigher focusing $w\sim 1$~$\upmu$m and longer pulses $\tau\sim 10$ns, or different material choices, this condition could be deliberately violated, but we do not consider such cases here.

Following the standard formulation~\cite{li2009photoacoustic}, we employ the scalar wave approximation, which adequately describes the predominantly longitudinal acoustic waves generated in the LIAD process. This linear model has formal limitations at high fluence where local temperatures exceed material thresholds; we discuss this further in the conclusions. We divide the standard photoacoustic equation by the acoustic impedance $Z = \rho v$ to obtain an equation for the particle velocity $u$ rather than pressure:
\begin{equation}\label{eq:waveeqn}
\left[\nabla^2 - v^{-2}\partial_t^2\right]u(\mathbf{r},t) = S(\mathbf{r},t) = -\frac{1}{Z}\frac{\alpha}{C_p}\partial_t H(\mathbf{r},t),
\end{equation}
where $S$ is the source term determined by the heating function $H$, the thermal expansion coefficient $\alpha$, and the specific heat capacity at constant pressure $C_p$.

The acoustic disturbance is found by convolving the source term with the scalar Green's function:
\begin{equation}
u(\mathbf{r},t) = \iint_\Omega G(\mathbf{r}-\mathbf{r}',t-t') S(\mathbf{r}',t') \, d\mathbf{r}' dt',
\end{equation}
where $G(\mathbf{r},t) = \delta(t-|\mathbf{r}|/v)/4\pi |\mathbf{r}|$ and $v$ is the speed of sound. The Dirac delta function allows the time integral to be evaluated analytically, yielding:
\begin{equation}
u(\mathbf{r},t) = \int_\Omega \frac{1}{4\pi|\mathbf{r}-\mathbf{r}'|}S(\mathbf{r}',t-|\mathbf{r}-\mathbf{r}'|/v)\, d\mathbf{r}'.
\end{equation}

\subsection{Laser pulse model}

We model a Gaussian laser pulse propagating from vacuum ($z<0$) in the $+z$ direction, incident on the material surface at $z=0$. The homogeneous substrate occupies $z>0$ with penetration depth $\xi$. The heating function combines the temporal Gaussian envelope with exponential absorption:
\begin{equation}
H(\mathbf{r},t) = H_0 e^{-(x^2+y^2)/2w^2}e^{-z/\xi}e^{-(t/\tau)^2},
\end{equation}
where the spatial profile follows the transverse beam intensity $e^{-(x^2+y^2)/2w^2}$ and the Beer-Lambert absorption law $e^{-z/\xi}$. The amplitude $H_0$ is determined by requiring that the total absorbed energy equals $\beta U_0$:
\begin{equation}
\iiint H(\mathbf{r},t) \, d\mathbf{r} \, dt = \beta U_0
\end{equation}
yielding $H_0 = \beta U_0/(2\pi^{3/2}w^2\xi\tau)$, where $U_0$ is the pulse energy, $\beta$ the absorption coefficient, $w$ the beam waist, and $\tau$ the pulse duration.

The corresponding source term, via Eq.~\ref{eq:waveeqn} becomes:
\begin{equation}
S(\mathbf{r},t) = -S_0 e^{-(x^2+y^2)/2w^2}e^{-z/\xi} \frac{t}{\tau} e^{-(t/\tau)^2}
\end{equation}
for $z > 0$, where $S_0 = \frac{1}{Z}\frac{\alpha}{C_p}\frac{2H_0}{\tau}$.

To account for acoustic reflection at the substrate surface $z=0$, we employ the method of images with the substitution $z \to |z|$ to extend the source distribution to $z < 0$ and integrate over $-\infty < z < +\infty$; the acoustic method of images has the same sign for the mirror source, contrasting with electromagnetism.

The particle-laden far surface can be placed at any depth $z > 0$ by evaluating the acoustic velocity or acceleration at that location. Acoustic reflections from the far surface will generate subsequent pulses that propagate within the finite-thickness substrate. However, we consider only the first arriving pulse: the short pulse duration prevents etalon-like interference effects, and the first pulse carries the largest amplitude since subsequent reflections experience cumulative diffraction losses.

\subsection{Dimensional reduction for on-axis detection}

For points directly above the center of the laser spot, cylindrical symmetry allows significant computational simplification. Setting the detection point at $\mathbf{r} = (0,0,z)$ the 3D integral becomes
\begin{align}\label{eq:onaxisdoubleintegral}
u(0,0,z,t) &= \frac{1}{2} \int_0^\infty \int_{-\infty}^\infty \frac{r'}{\sqrt{r'^2+(z-z')^2}} \nonumber \\
&\quad \times S(r',z',t-\sqrt{r'^2+(z-z')^2}/v) \, dr' dz'
\end{align}
which is easily computationally tractable. We note that time ($t/\tau$) and length scales appear as ratios (e.g. $x^2/w^2$), which provides freedom to choose natural units in the numerical calculations.

\subsection{Analytical performance scaling}

The treatment provides scaling laws without the need to compute the full integral.
The surface acceleration amplitude scales as
\begin{equation}
a_{\mathrm{max}} \propto \frac{S_0}{\tau} \propto \frac{\alpha \beta U_0}{\rho v C_p \tau^2 w^2 \xi}
\end{equation}
which reveals key dependencies which we now examine in turn: pulse duration, wavelength and material selection, and spot size.

\subsubsection{Pulse duration optimization}

Acceleration scales as $\tau^{-2}$, meaning shorter pulses are desirable. Physical constraints limit this optimization:

\textit{Lower bound}: Pulse duration has a lower bound: $\tau \gtrsim 10$~ps, where the photoacoustic approximation breaks down due to electron-phonon coupling~\cite{allen1987theory,groeneveld1995femtosecond}.

\textit{Upper bound}: At $\tau \gtrsim 1$~$\upmu$s, thermal diffusion during the pulse breaks the thermal confinement assumption~\cite{xu2006photoacoustic}.

Passively Q-switched lasers achieve sub-nanosecond durations near the theoretical optimum within the photoacoustic regime.

\subsubsection{Wavelength and absorption considerations}

Traditional LIAD uses 532~nm (frequency-doubled Nd:YAG), achieving modestly better aluminum absorption compared with the fundamental at 1064~nm ($\beta \sim 0.08$ vs $0.05$). However, operating without doubling may be preferable as the lower absorption is offset by avoiding losses inherent in second harmonic generation, which typically reduce pulse energy by 50\% or more.

The material-dependent optical and thermomechanical properties suggest a figure of merit
\begin{equation}
\mathrm{FOM} \propto \frac{\alpha \beta}{\rho v C_p \xi}.
\end{equation}
Aluminum is the traditional substrate choice, but tungsten and titanium may be preferable due to higher thermal expansion and superior infrared optical properties. We examine these material choices quantitatively in Sec.~\ref{sec:case_studies}, using optical constants from Table~\ref{tab:optical_props}.

\subsubsection{Optimal spot size and acoustic diffraction}

Smaller beam waists increase energy density ($w^{-2}$ scaling), but also increase the divergence of the resulting acoustic wave as it propagates through the substrate thickness $d$.
Finding parameters which optimise acceleration at the far surface must include this diffraction.
If the acoustic wave is to remain approximately collimated over the thickness $d$ then, treating this acoustic wave as a Gaussian beam with a characteristic Rayleigh range, we find
\begin{equation}
w \gtrsim \sqrt{\frac{\lambda_a d}{\pi}} \sim \sqrt{v\tau d}
\end{equation}
where $\lambda_a = v\tau$ is the acoustic wavelength.

\subsection{Acoustic wave propagation dynamics}

\begin{figure}[t]
\centering
\includegraphics[width=\columnwidth]{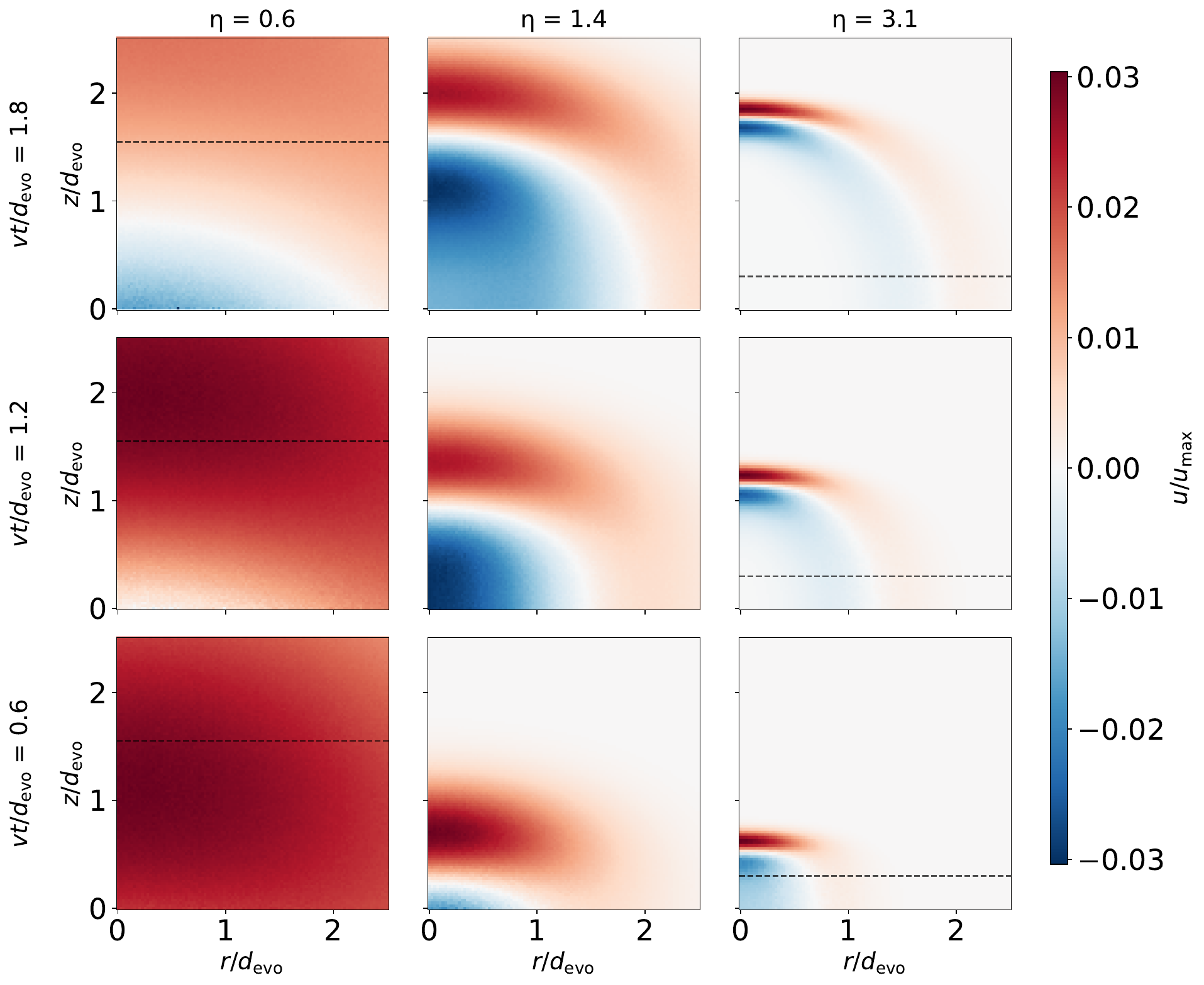}
\caption{Acoustic wave propagation and diffraction through metal substrates. Cross-sectional views show the acoustic velocity $u(r,z,t)$ at time instances (lower rows are earlier times) during propagation from laser-heated front surface ($z=0$) towards the particle-laden surface ($z>0$). The dimensionless focusing parameter $\eta=w/(v\tau)=d_{\mathrm{evo}}/w$ is discussed in the text. Left and right columns correspond to the systems of Nikkhou et al.~($\eta = 0.6$) and Bykov et al.~($\eta = 3.1$) respectively; horizontal dashed lines indicate their substrate thicknesses ($d/d_{\mathrm{evo}} = 1.55$ and $0.30$ respectively), showing that the Nikkhou pulse has largely evolved while the Bykov pulse is still evolving.}
\label{fig:diffraction}
\end{figure}

The acoustic wave exhibits two key behaviors: first, the wave diffracts laterally as it propagates (Figure~\ref{fig:diffraction}); second, and less obviously, the on-axis temporal pulse shape evolves during propagation because it arises from an extended source.

\subsubsection{Characteristic pulse shape evolution distance}

The photoacoustic wave begins with a spatial profile determined by the exponential Beer-Lambert absorption, $\exp(-z/\xi)$, which evolves as it propagates.
Two physical processes compete to set the evolution length scale: the optical skin depth $\xi$, and the acoustic diffraction (Fresnel) length $d_{\mathrm{evo}}=w^2/(v\tau)$. The connection to pulse duration arises because, together with the speed of sound, this sets the acoustic wavelength.

For typical metal substrates and laser parameters, the skin depth $\xi \sim 10$~nm is negligible compared to diffraction effects $d_{\mathrm{evo}} \sim 10{-}1000$~$\upmu$m; 
table~\ref{tab:laser_params} shows values for various experimental configurations.
Tightly focused systems (Nikkhou: 17~$\upmu$m waist) achieve $d_{\mathrm{evo}} \sim 11$~$\upmu$m, where pulse shape stabilizes quickly relative to typical substrate thicknesses. Large-waist systems exhibit slow evolution: Northup ($d_{\mathrm{evo}} \sim 60$~$\upmu$m) shows moderate thickness dependence, while proposed shorter pulse systems (passive Q-switched: $d_{\mathrm{evo}} \sim 1500$~$\upmu$m) continue evolving throughout typical substrate thicknesses.

These two distinct regimes may have complementary use cases: systems with $d_{\mathrm{evo}} \ll d$ give consistent profiles independent of substrate thickness, while $d_{\mathrm{evo}} \gtrsim d$ provides a potentially useful thickness-dependent behaviour.

\subsubsection{Analytical solution in the thin-skin limit}

Since $\xi \ll d_{\mathrm{evo}}$ for all practical LIAD systems, the exponential Beer-Lambert absorption $\exp(-|z'|/\xi)$ can be approximated by delta function $\xi\,\delta(z')$ in the source term. The double integral Eq.~\ref{eq:onaxisdoubleintegral} then reduces to a single integral over the radius, which can be evaluated analytically (Appendix~\ref{app:analytical}).

The analytical solution reveals that, once rescaled to natural units $(z/d_{\mathrm{evo}}, vt/d_{\mathrm{evo}})$ there remains a shape dependence on a single dimensionless parameter:
\begin{equation}
\eta = \frac{w}{v\tau} = \frac{d_{\mathrm{evo}}}{w}
\label{eq:eta}
\end{equation}
which compares the beam waist to the acoustic wavelength $v\tau$. This parameter characterizes the diffraction regime: small $\eta$ corresponds to tight focusing with strong diffraction, while large $\eta$ indicates collimated propagation.

\begin{figure}[t]
\centering
\includegraphics[width=\columnwidth]{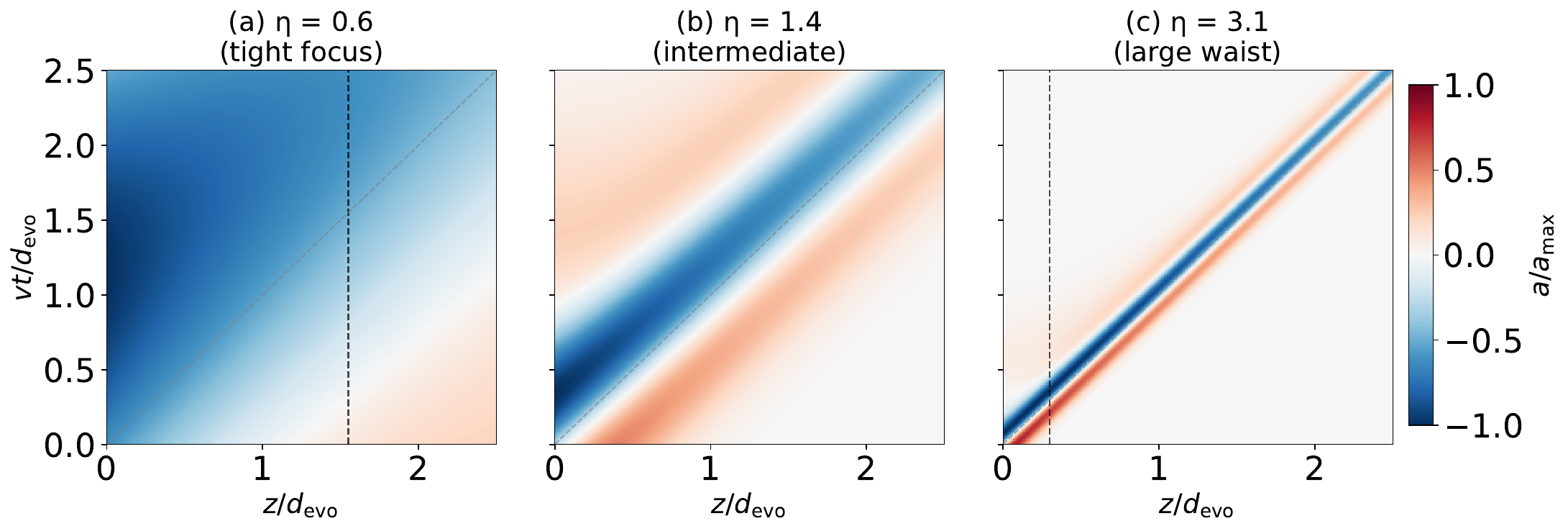}
\caption{Acoustic pulse evolution in the thin-skin limit for different diffraction regimes. Normalized acceleration $a/a_{\mathrm{nat}}$ (scaled to peak) as a function of dimensionless coordinates $z/d_{\mathrm{evo}}$ and $vt/d_{\mathrm{evo}}$ for (a) $\eta = 0.6$ (Nikkhou et al., $d/d_{\mathrm{evo}} = 1.55$), (b) $\eta = 1.4$ (intermediate), and (c) $\eta = 3.1$ (Bykov et al., $d/d_{\mathrm{evo}} = 0.30$). The dimensionless parameter $\eta = w/(v\tau)$ characterizes the diffraction regime and determines the evolution shape. Vertical dashed lines mark these substrate thicknesses; gray diagonal line shows the characteristic ($vt = z$). Computed from analytical solution Eq.~\ref{eq:analytical_acceleration}.}
\label{fig:spacetime}
\end{figure}

Figure~\ref{fig:spacetime} shows the acoustic pulse evolution for different values of $\eta$. The evolution distance $d_{\mathrm{evo}}$ sets the length scale, while $\eta$ determines the shape. Systems with identical $d_{\mathrm{evo}}$ but different $\eta$ exhibit qualitatively different evolution patterns. For typical LIAD parameters (Table~\ref{tab:laser_params}), $\eta$ ranges from $\sim 0.6$ (Nikkhou: tight focus) to $\sim 30$ (passive Q-switched: short pulse duration).

\section{Case studies and parameter optimization}
\label{sec:case_studies}

LIAD surface accelerations are difficult to measure directly. Piezoelectric launchers provide a complementary and more easily estimated threshold acceleration $\sim 10^8$~g~$\sim1$~nm$/$ns$^2$ needed for 100~nm particles~\cite{khodaee2022dry}. We use these values as design targets.

We validate our framework using representative LIAD implementations from literature, and then explore optimization opportunities with commercial laser systems. Tables~\ref{tab:laser_params}--\ref{tab:optical_props} provide the independent physical parameters governing LIAD performance.
The diffraction parameter $\eta = w/(v\tau)$ varies by more than an order of magnitude across these systems.

\begin{table*}[t]
\centering
\caption{Laser system parameters for literature examples (Bykov et al., 2019; Nikkhou et al., 2021) and theoretically informed choices, parameters of which are based on commercially available laser systems. SHG denotes second harmonic generation (frequency doubling). Passive Q-switched systems typically provide excellent beam quality (M$^2 \lesssim 1.2$) and can therefore focus more tightly, but optimal spot size is limited by substrate thickness to avoid acoustic diffraction. The evolution distance $d_{\mathrm{evo}} = w^2/(v\tau)$ and diffraction parameter $\eta = w/(v\tau)$ are calculated using aluminum substrate properties ($v = 6420$~m/s).}
\label{tab:laser_params}
\begin{tabular}{lccccccc}
\hline
Case & $\lambda$ [nm] & $\tau$ [ns] & $U$ [$\upmu$J] & $w$ [$\upmu$m] & M$^2$ & $d_{\mathrm{evo}}$ [$\upmu$m] & $\eta$ \\
\hline
\multicolumn{8}{l}{\textit{Literature examples:}} \\
Bykov et al.~\cite{bykov2019direct} (SHG Nd:YAG) & 532 & 5.0 & 3000 & 100 & --- & 312 & 3.1 \\
Nikkhou et al.~\cite{nikkhou2021direct} (SHG Nd:YAG) & 532 & 4.6 & 3000 & 17 & --- & 10 & 0.6 \\
\hline
\multicolumn{8}{l}{\textit{Theoretically informed choices:}} \\
Passive Q-switched Nd:YAG & 1064 & 0.6 & 100 & 100\rlap{$^*$} & $<1.2$ & 2600 & 26 \\
Pulsed diode laser & 1030 & 1.3 & 100 & 100 & $>5$ & 1200 & 12 \\
\hline
\multicolumn{8}{l}{\footnotesize $^*$Can be focused much tighter; limited by acoustic diffraction} \\
\end{tabular}
\end{table*}

\begin{table}[h]
\centering
\caption{Substrate material properties (wavelength-independent). Values from CRC Handbook~\cite{haynes2016crc}.}
\label{tab:material_props}
\begin{tabular}{lD{.}{.}{5}D{.}{.}{4}D{.}{.}{2.1}D{.}{.}{3}}
\hline
Material & \multicolumn{1}{c}{$\rho$ [kg/m$^3$]} & \multicolumn{1}{c}{$v$ [m/s]} & \multicolumn{1}{c}{$\alpha$ [10$^{-6}$/K]} & \multicolumn{1}{c}{$C_p$ [J/kg$\cdot$K]} \\
\hline
Al & 2700 & 6420 & 23.1 & 897 \\
StSt & 7900 & 5790 & 17.3 & 483 \\
Ti & 4506 & 6070 & 8.6 & 524 \\
W & 19300 & 5220 & 4.5 & 132 \\
\hline
\end{tabular}
\end{table}

\begin{table}[h]
\centering
\caption{Wavelength-dependent optical properties: skin depth $\xi$ and absorption coefficient $\beta$. Values calculated from literature refractive indices~\cite{rakic1995algorithm,johnson1974optical,ordal1988optical}.}
\label{tab:optical_props}
\begin{tabular}{lD{.}{.}{2.1}D{.}{.}{2.1}D{.}{.}{2.1}D{.}{.}{2.1}}
\hline
Wavelength & \multicolumn{2}{c}{532 nm} & \multicolumn{2}{c}{1064 nm} \\
Material & \multicolumn{1}{c}{$\beta$ [\%]} & \multicolumn{1}{c}{$\xi$ [nm]} & \multicolumn{1}{c}{$\beta$ [\%]} & \multicolumn{1}{c}{$\xi$ [nm]} \\
\hline
Al & 8 & 6.6 & 5 & 8.3 \\
StSt & 49 & 14.5 & 37 & 21.2 \\
Ti & 42 & 12.6 & 38 & 21.1 \\
W & 48 & 14.6 & 40 & 22.4 \\
\hline
\end{tabular}
\end{table}

\begin{figure}[t]
\centering
\includegraphics[width=\columnwidth]{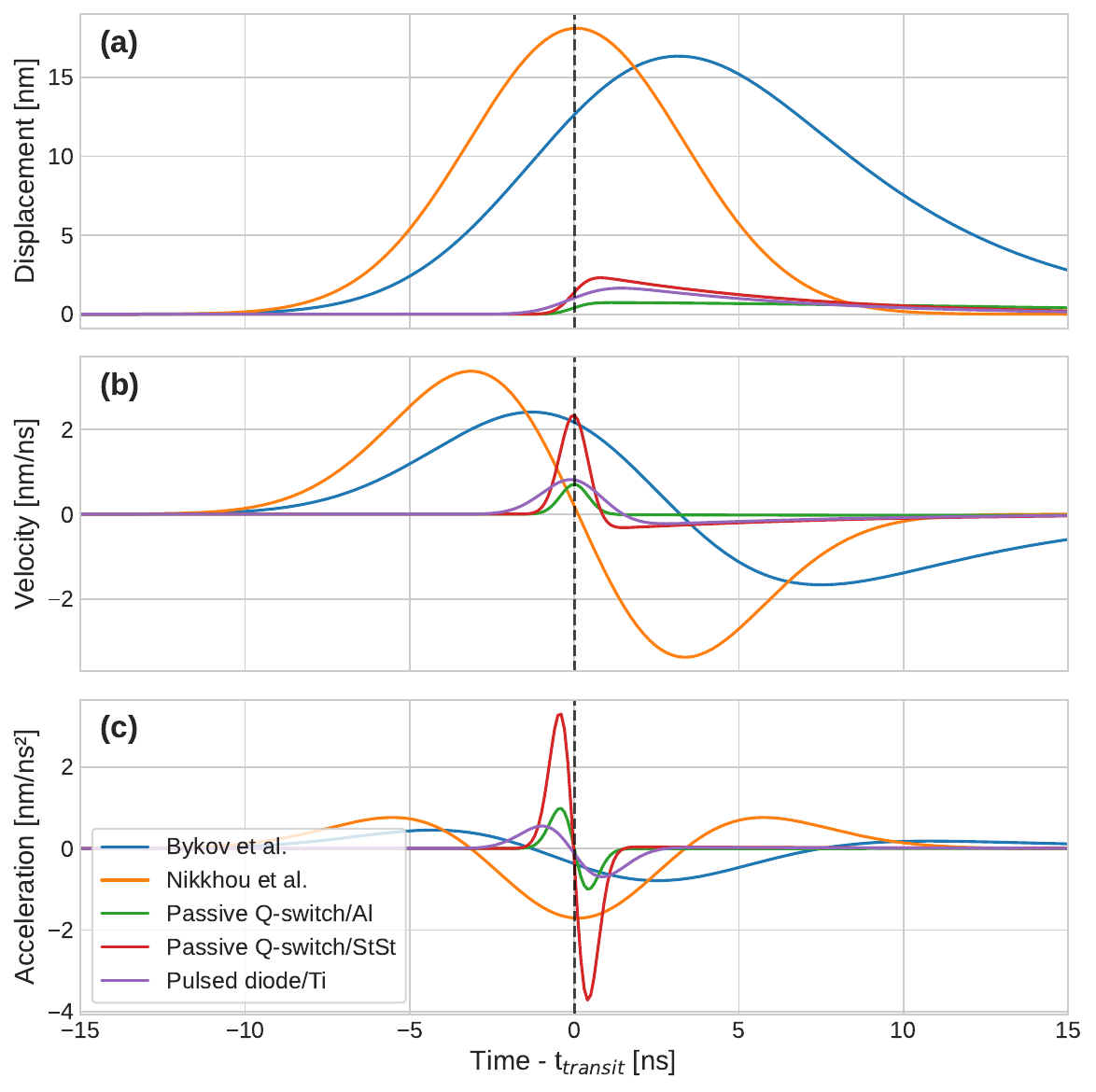}
\caption{Photoacoustic response comparison for literature (Bykov et al., 2019; Nikkhou et al., 2021) and commercial systems: (a) displacement, (b) velocity, and (c) acceleration. The time axis centers on acoustic transit time (thickness/v) to ease direct comparison. The surface is confirmed to return to zero displacement after the pulse. Passive Q-switched systems perform competitively, insofar as they achieve comparable or larger surface accelerations, despite lower pulse energies. This model predicts that aluminium substrates show lower peak accelerations than optimized materials like stainless steel or titanium.}
\label{fig:comparison}
\end{figure}

The main results of this section are presented in Figure~\ref{fig:comparison}.
While existing systems demonstrate large surface \emph{displacements}, resulting from their comparitively large pulse energy, shorter pulses can more than compensate for their lower pulse energy by the shortness and consequently larger \emph{acceleration}.  All systems exceed the threshold $1~$nm$/$ns$^2$ suggested by piezo-based systems.

The system of Nikkhou has a short $d_{\textrm{evo}}$ and has therefore reached a steady-state propagating wave, while that of Bykov has a larger $d_{\textrm{evo}}$ and is therefore still evolving in shape when it reaches the far surface of the substrate; this is also the case for the two proposed systems, for typical substrate thicknesses.

We note that the displacement traces provide a consistency check: the source term $S(\mathbf{r},t) \propto (t/\tau) \exp(-(t/\tau)^2)$ is an odd function in time, and hence $\int_{-\infty}^{\infty} u(\mathbf{r},t) dt = 0$, which corresponds to the surface returning to its original position after the acoustic wave passes. 

\subsection{Optical properties}

Table~\ref{tab:optical_props} values are calculated from complex refractive index $\tilde{n} = n + i\kappa$ using
\begin{equation}
\xi = \frac{\lambda_0}{4\pi \kappa}, \quad
R = \frac{(n-1)^2 + \kappa^2}{(n+1)^2 + \kappa^2}, \quad
\beta = 1 - R
\end{equation}
where $\lambda_0$ is vacuum wavelength. Complex refractive index values at 532~nm and 1064~nm were extracted from the refractiveindex.info database~\cite{polyanskiy2024refractiveindexinfo}: aluminum from Rakić~\cite{rakic1995algorithm}, titanium and iron (stainless steel proxy) from Johnson \& Christy~\cite{johnson1974optical}, and tungsten from Ordal et al.~\cite{ordal1988optical}.

Optical constants for metals exhibit substantial variation across the literature due to surface preparation, oxidation, and measurement techniques. However, the key scaling relationships (acceleration $\propto \tau^{-2}$, optimal spot size $w \sim \sqrt{v\tau d}$) depend primarily on fundamental material properties (density, sound velocity) rather than precise optical constants. The skin depth $\xi$ affects the spatial distribution of the heat source, but for substrate thicknesses $\sim$100~µm, acoustic diffraction limitations dominate over optical penetration depth effects.

\section{Conclusions}

We have developed a photoacoustic model for systematic optimization of laser-induced acoustic desorption.
The model identifies key scaling relationships (surface acceleration $\propto \tau^{-2}$ and optimal spot size $w \gtrsim \sqrt{v\tau d}$) which enables parameter selection based on material properties and performance requirements rather than empirical recipes.

Validation against published experimental implementations shows good agreement, with predicted accelerations consistent with reported successful implementations of nanoparticle desorption. These implementations operate in qualitatively different regimes, which our model provides a means to describe. The model further reveals that compact laser systems with sub-nanosecond pulse durations can achieve competitive performance despite significantly lower pulse energies than traditional laboratory implementations. Material figures of merit suggest alternatives to standard aluminum substrates that may improve performance at infrared wavelengths.

We note that the thin skin limit, which allows analytical treatment of the on-axis case, will not hold in all systems. For example, a silicon substrate at 1064~nm has a skin depth $\xi\sim 10\,\upmu$m which is much larger than for metals at this wavelength. Treatment of such scenarios requires the full numerical integral, as we have done for all case studies in the present work.

The photoacoustic model has formal limitations at typical LIAD parameters. Estimating the peak temperature rise within the optical skin depth for a 3~mJ pulse absorbed in volume $\sim \pi w^2 \xi$ gives values far exceeding the material melting point, indicating operation well outside the linear thermoelastic regime. Ablation has been observed in LIAD implementations of this type. Despite these limitations, the model successfully predicts acoustic amplitudes consistent with published implementations (Fig.~\ref{fig:comparison}). The acoustic wave is generated during the initial rapid heating, before thermal diffusion or phase change significantly alter the dynamics. The scaling relationships ($a \propto \tau^{-2}$, optimal $w \sim \sqrt{v\tau d}$) depend on ratios of material properties rather than absolute temperature rises, which may explain their continued validity. The model is best understood as capturing relative performance across parameter regimes rather than absolute local thermodynamics.

Acoustic attenuation provides an additional consideration for short-pulse systems. At the 100--300~MHz frequencies corresponding to nanosecond pulses, attenuation in aluminum substrates is modest over typical thicknesses~\cite{truell1969ultrasonic}. Sub-nanosecond pulses approach gigahertz frequencies where attenuation increases, potentially reducing the $\tau^{-2}$ advantage at the shortest pulse durations.

\section*{Data Availability Statement}

The simulation code that supports the findings of this article is openly available~\cite{edmonds2025photoacoustic-code}.

\section*{Acknowledgments}

This work was supported by the UK Space Agency under the Enabling Technologies Programme, project ETP1-025 ``LOTIS'', and by the Engineering and Physical Sciences Research Council [EP/X524979/1] with the Defence Science and Technology Laboratory (DSTL) under iCASE voucher number 220010.

\bibliography{bibliography}

\onecolumngrid
\appendix

\section{Analytical solution in the thin-skin limit}
\label{app:analytical}

For practical LIAD systems, the optical skin depth $\xi \sim 10$~nm is negligible compared to both the beam waist $w \sim 10{-}100$~$\upmu$m and the acoustic wavelength $v\tau \sim 10{-}100$~$\upmu$m. In this thin-skin limit, the exponential Beer-Lambert absorption $\exp(-|z'|/\xi)$ can be replaced by a Dirac delta function in the source term.

The method of images (Sec.~2.2) extends the source to negative $z'$ with $\exp(-|z'|/\xi)$. Integrating over $-\infty < z' < \infty$ gives $\int_{-\infty}^{\infty} \exp(-|z'|/\xi) dz' = 2\xi$, introducing a factor of 2 from the acoustic reflection at the surface. The source term becomes:
\begin{equation}
\label{eq:source_thinskin}
S(r',z',t) = -S_0 e^{-r'^2/(2w^2)} 2\xi\,\delta(z') \frac{t}{\tau} e^{-(t/\tau)^2}
\end{equation}
where we use cylindrical coordinates with $r' = \sqrt{x'^2+y'^2}$.

Substituting Eq.~\ref{eq:source_thinskin} into the on-axis integral (Eq.~\ref{eq:onaxisdoubleintegral}) and exploiting the sifting property of the Dirac delta function, the $z'$ integral collapses to evaluation at $z'=0$:
\begin{align}
u(z,t) = -\xi S_0 \int_0^\infty \frac{r'}{\sqrt{r'^2+z^2}} e^{-r'^2/(2w^2)}\times \frac{t-R/v}{\tau} \exp\left[-\left(\frac{t-R/v}{\tau}\right)^2\right] dr'
\label{eq:thinskin_integral}
\end{align}
where $R = \sqrt{r'^2+z^2}$.

We introduce the evolution distance $d_{\mathrm{evo}} = w^2/(v\tau)$ as the natural length scale.
We define dimensionless coordinates
\begin{equation}
z_{\mathrm{nat}} = \frac{z}{d_{\mathrm{evo}}}, \quad t_{\mathrm{nat}} = \frac{t}{d_{\mathrm{evo}}/v}
\label{eq:normalized_coords}
\end{equation}
and the dimensionless parameter
\begin{equation}
\eta = \frac{w}{v\tau} = \frac{d_{\mathrm{evo}}}{w}
\label{eq:eta_def}
\end{equation}
which compares the beam waist to the acoustic wavelength.
The normalized velocity is
\begin{equation}
u_{\mathrm{nat}}(z_{\mathrm{nat}},t_{\mathrm{nat}}) = \frac{u(z,t)}{S_0 \xi d_{\mathrm{evo}}}.
\end{equation}

For brevity in the coming equations we also define
\begin{equation}
\mu = 1 + 2\eta^2, \quad \Delta = z_{\mathrm{nat}} - t_{\mathrm{nat}}
\label{eq:mu_delta_def}
\end{equation}
where $\Delta$ is the retarded coordinate measuring distance from the acoustic wavefront.

The integral Eq.~\ref{eq:thinskin_integral} can be evaluated analytically, yielding:
\begin{align}
u_{\mathrm{nat}}(z_{\mathrm{nat}}, t_{\mathrm{nat}})
&=
\frac{1}{2\mu^{3/2}}
\exp\!\left[
  -\eta^{4}
  \left(
    z_{\mathrm{nat}}^{2}
    + \frac{t_{\mathrm{nat}}^{2}(\mu+1)}{\mu}
  \right)
\right]
\nonumber\\[1em]
&\quad\times
\biggl[
2\sqrt{\mu}\exp\!\left(
     \frac{t_{\mathrm{nat}}\eta^{4}(t_{\mathrm{nat}} + 2\mu z_{\mathrm{nat}})}{\mu}
   \right)
\nonumber\\[0.5em]
&\qquad\qquad
- \sqrt{2\pi}\,t_{\mathrm{nat}}\eta
    \exp\!\left(
      \frac{\mu\eta^{2}z_{\mathrm{nat}}^{2}}{2}
      +\eta^{4}t_{\mathrm{nat}}^{2}
    \right)
    \mathrm{erfc}\!\left(
      \frac{\eta(\mu\Delta + t_{\mathrm{nat}})}{\sqrt{2\mu}}
    \right)
\biggr]
\label{eq:analytical_solution}
\end{align}
where $\mathrm{erfc}(x) = 1 - \mathrm{erf}(x)$ is the complementary error function.

The normalized acceleration is obtained by taking the time derivative:
\begin{align}
a_{\mathrm{nat}}(z_{\mathrm{nat}}, t_{\mathrm{nat}})
&=
\frac{\partial u_{\mathrm{nat}}}{\partial t_{\mathrm{nat}}}\nonumber\\
&=
\frac{\eta}{2\mu^{3}}
\exp\!\left(
  2\eta^{4}t_{\mathrm{nat}}\Delta
  -\frac{\eta^{2}(2\mu-1)z_{\mathrm{nat}}^{2}}{2}
\right)
\nonumber\\[1em]
&\quad\times
\biggl[
4\mu\eta^{3}(\mu\Delta - t_{\mathrm{nat}})
\exp\!\left(
  \eta^{4}t_{\mathrm{nat}}^{2}
  +\frac{\mu\eta^{2}z_{\mathrm{nat}}^{2}}{2}
\right)
\nonumber\\[0.5em]
&\qquad\qquad
+ \sqrt{2\pi\mu}\left(2\eta^{4}t_{\mathrm{nat}}^{2}-\mu\right)
    \exp\!\left(
      \frac{\mu\eta^{2}z_{\mathrm{nat}}^{2}}{2}
      +\eta^{4}t_{\mathrm{nat}}^{2}
      +\frac{\eta^{2}(\mu\Delta + t_{\mathrm{nat}})^{2}}{2\mu}
    \right)
\nonumber\\[0.5em]
&\qquad\qquad\quad
    \times\mathrm{erfc}\!\left(
      \frac{\eta(\mu\Delta + t_{\mathrm{nat}})}{\sqrt{2\mu}}
    \right)
\biggr].
\label{eq:analytical_acceleration}
\end{align}

\end{document}